\documentclass[conference]{IEEEtran}
\usepackage{graphicx}  
\usepackage{amsmath,amssymb,amstext,amsbsy}
\begin{document}
\title{Waiting time analysis of foreign currency exchange rates: 
Beyond the renewal-reward theorem}
\author{\authorblockN{Naoya Sazuka\authorrefmark{1} and 
Jun-ichi Inoue\authorrefmark{2}}
\authorblockA{\authorrefmark{1}Sony Corporation, 
4-10-18 Takanawa Minato-ku, 
Tokyo 108-0074, Japan \\
Email: Naoya.Sazuka@jp.sony.com \\
\authorrefmark{2}Complex Systems Engineering, 
Graduate School of Information Science and 
Technology \\
Hokkaido University, 
N14-W9, Kita-ku, Sapporo 060-0814, Japan \\
Email: j$\underline{\,\,\,}$inoue@complex.eng.hokudai.ac.jp}
}
\markboth{Journal of \LaTeX\ Class Files,~Vol.~1, No.~11,~November~2002}{Shell \MakeLowercase{\textit{et al.}}: Bare Demo of IEEEtran.cls for Journals}
\maketitle
\begin{abstract}
We evaluate the average waiting time
between observing the price of financial markets
and the next price change, 
especially in an on-line foreign exchange trading service
for individual customers
via the internet.
Basic technical idea 
of our present work is dependent on 
the so-called renewal-reward theorem. 
Assuming that stochastic 
processes of the market price changes 
could be regarded as a renewal process, 
we use the theorem to calculate the average waiting time 
of the process. 
In the conventional 
derivation of the theorem, 
it is apparently 
hard to evaluate the higher order moments of 
the waiting time. 
To overcome this type of 
difficulties, we attempt to derive 
the waiting time distribution 
$\Omega (s)$ directly 
for arbitrary time interval 
distribution 
(first passage time distribution) 
of the stochastic process 
$P_{W}(\tau)$ and observation time distribution 
$P_{O}(t)$ of customers. 
Our analysis enables us to evaluate not only the first moment 
(the average waiting time) but also any order of the 
higher moments of 
the waiting time. 
Moreover, in our formalism, 
it is possible to 
model the observation 
of the price on the internet by the customers 
in terms of the observation time 
distribution $P_{O}(t)$. 
We apply our analysis to the stochastic process 
of the on-line foreign exchange rate
for individual customers 
from the Sony bank 
and compare the moments 
with the empirical data analysis. 
\end{abstract}

\IEEEpeerreviewmaketitle
\section{Introduction}
\label{sec:Intro}
Fluctuation has an important role in lots of 
phenomena appearing in our real world. 
For instance, 
a kind of magnetic alloy possesses 
magnetism in low temperature, whereas 
in high temperature, 
it losses the magnetism 
due to thermal fluctuation 
acting on each spin (a tiny magnet in atomic scale length). 
Thus, the large system undergoes 
a phase transition at some critical temperature and 
the transition occurs 
due to cooperative behavior of huge number of 
spins, to put it other way, 
due to competition between 
thermal fluctuation and exchange 
interaction between spin pairs 
making them point to the same direction 
\cite{Stanley}. 

To understand these kinds of macroscopic 
properties or collective behavior of the system 
from the microscopic point of view, 
statistical mechanics provides a good tool. 
Actually, 
statistical mechanics has been 
applied to various research subjects in which 
fluctuation is an essential key point, 
such as information 
processing \cite{Nishi00} or 
economics including financial markets and 
game theory \cite{Coolen}. 
Especially, 
application of 
statistical-mechanical 
tools to economics, 
data analysis of financial markets --- 
what we call {\it econophysics} --- 
is one of the most developing research fields 
\cite{Mantegna2000,Bouchaud,Voit}.
Financial data have attracted a lot 
of attentions of physicists as informative materials to 
investigate the macroscopic behavior of the markets from 
the microscopic statistical properties \cite{Mantegna2000,Bouchaud,Voit}. 
Some of these studies are restricted to 
the stochastic variables of the price changes 
(returns) and most of them is specified by a key word, that is 
to say, {\it fat tails} of the 
distributions \cite{Mantegna2000}. However, 
the distribution of time intervals 
also might have important information about the 
markets and it is worth while for us to investigate these 
properties extensively 
\cite{Engle98,Mainardi,Raberto,Scalas,Kaizoji,Scalas06}. 

Such kinds of fluctuation 
in time intervals between events 
are not special phenomena in price changes 
in financial markets but very common in science. 
In fact, it is well-known that spike train 
of a single neuron in real brain 
is time series in which the difference between successive 
two spikes is not constant but fluctuated. 
This stochastic process specified by the so-called 
Inter-Spike Intervals (ISI) 
is one of such examples \cite{Tuckwell,Gerstner}. 
The average of 
the ISI is about a few milli-second 
and the distribution of the intervals is 
well-described by the {\it Gamma distribution} \cite{Gerstner}. 

On the other hand, 
in financial markets, 
for instance, the time intervals of 
two consecutive transactions 
of BUND futures (BUND is the German word for bond)
and BTP futures
(BTP is the middle and long term Italian Government bonds 
with fixed interest rates)
traded at LIFFE 
(LIFFE stands for London International Financial Futures and Options Exchange) 
are $\sim 10$ seconds 
and are well-fitted by the so-called {\it Mittag-Leffler 
function} \cite{Mainardi,Raberto,Scalas}. 
The Mittag-Leffler 
function behaves as a stretched exponential 
distribution for 
short time interval regime, 
whereas for the long time interval regime, 
the function has a power-law tails. 
Thus, the behavior of the distribution 
described by the Mittag-Leffler function 
is changed from the stretched 
exponential to the power-law at some critical point 
\cite{Gorenflo}. 
However, it is non-trivial to 
confirm if the Mittag-Leffler function supports 
any other kind of market data, for example, 
the market data filtered by some rate window. 

As such market data, 
the Sony bank USD/JPY exchange rate \cite{Sony},
which is the rate
for individual customers
of the Sony bank
in their on-line foreign exchange trading service, 
is a good example to be checked by 
the Mittag-Leffler function.
Actually, our preliminary 
results imply that 
the Mittag-Leffler function 
does not support the Sony bank rate \cite{SazukaInoue2006}. 
The Sony bank rate has 
$\sim 20$ minutes \cite{Sazuka} 
as the average time interval 
which is extremely longer than the other 
market rate as the BUND future. 
This is because the Sony back rate 
can be regarded as the so-called {\it first passage process} 
\cite{Redner,Kappen,Gardiner,Risken,Simonsen,Kurihara}
of the raw market data. 
\begin{table}
\caption{\label{tab:table0}
Typical three examples with fluctuation between 
the events.
}
\begin{tabular}{|c||c|c|c|}
\hline 
\mbox{} & 
ISI & BUND future & Sony bank rate  \\
\hline
Average time interval 
& $\sim 3$ [ms]  & $\sim 10$ [s] & $\sim 20$ [min] \\
\hline
PDF & Gamma & 
Mittag-Leffler & Weibull \\
\hline
\end{tabular}
\end{table}
In Table \ref{tab:table0}, we list 
the average time intervals and 
the probability distribution function 
(PDF) that describes the data 
with fluctuation between the events for 
typical three examples, 
namely, 
the ISI, the BUND future and 
the Sony bank rate. 
From this table, 
an important question might be 
arisen. 
Namely, how long do the customers 
of the Sony bank should wait
between observing the price 
and the next price change? 
This type of 
question is never occurred 
in the case of the ISI or 
the BUND future because 
the average time intervals are too short to 
evaluate such informative measure. 

Obviously, for the customers, an important 
(relevant) quantity is sometimes a waiting time rather than a time interval  
between the rate changes. The waiting time we mentioned here 
is defined by the time for the customers to wait until the 
next price change since they try to observe it on 
the World Wide Web for example \cite{Sony}. 
If the sequence of the time intervals has some correlations and 
the customers observe the rate at random on the time axis, 
the distribution of the waiting time is no longer identical 
to the distribution of the time intervals. 
In the previous studies \cite{Sazuka0,Sazuka2,InoueSazuka2006}, 
we assumed that the time intervals of 
the Sony bank USD/JPY exchange rates might follow a {\it Weibull 
distribution} and evaluated the average waiting time by means of 
the renewal-reward theorem \cite{Tijms,Oishi}. 
However, the conventional renewal-reward theorem is 
restricted to the case in which the customers observe the rate at 
random on the time axis and it is hard to extend the theorem 
to the situation in which the time for the customers to 
observe the rates obeys some arbitrary distributions. 

To make these problems and difficulties clear, in this paper, 
we introduce a different way from the renewal-reward theorem
to evaluate the higher-order 
moments of the waiting time for 
arbitrary time interval distribution of the price changes 
and observation time distribution by 
directly deriving the waiting time distribution. 
We first show that the result of the renewal-reward theorem 
\cite{InoueSazuka2006} 
is recovered from our new formalism. Then, it becomes clear that 
our formulation is more general than the 
conventional renewal-reward theorem. 
As an advantage of our approach over the renewal-reward theorem, 
we can evaluate the higher-order moments of the waiting time, 
and moreover, it becomes possible to consider various 
situations in which the customers observe the rate according to 
arbitrary distribution of time. 

This paper is organized as follows. In the next section \ref{sec:SBR}, 
we introduce the Sony bank rate \cite{Sony} 
which is generated from 
the high-frequency foreign exchange market rate
via the rate window with width $2\epsilon$ yen 
($\epsilon=0.1$ yen for the Sony bank). 
In section \ref{sec:SumWorks}, 
we summarize a series of our previous 
studies related to the present paper.  
To understand the mechanism of the Sony bank rates as the 
first passage process 
\cite{Redner,Kappen} of the raw market rate, 
in 
section \ref{sec:GARCH}, 
we carry out 
the computer simulations by making use of the GARCH 
(Generalized AutoRegressive Conditional 
Heteroscedasicity) model 
\cite{Engle,Ballerslev,Franke} 
with discrete returns as 
the time series behind the Sony bank rates. 
The effect of the rate window on the currency 
exchange rates is revealed. 
In the next section \ref{sec:Deriv}, we explain our method 
to derive the waiting time distributions. We show that our treatment 
reproduces the result by the renewal-reward theorem. 
We also evaluate the deviation around the average waiting time for 
the Weibull first passage time distribution and uniform observation 
time distribution. We find that the resultant standard deviation is 
the same order as the average waiting time. 
We test our analysis for several cases of the observation time 
distributions and calculate the higher-order moments. 
The last section \ref{sec:Summary} is concluding remarks. 
\section{The Sony bank rates as a first passage process}
\label{sec:SBR}
\begin{figure}[ht]
\begin{center}
\includegraphics[width=8.5cm]{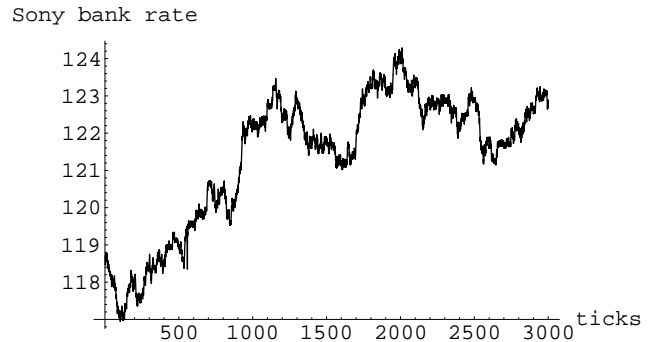}
\end{center}
\caption{\footnotesize 
Behavior of the Sony bank rates.}
\label{fig:fg_sony}
\end{figure}

\begin{figure}[ht]
\begin{center}
\includegraphics[width=8.5cm]{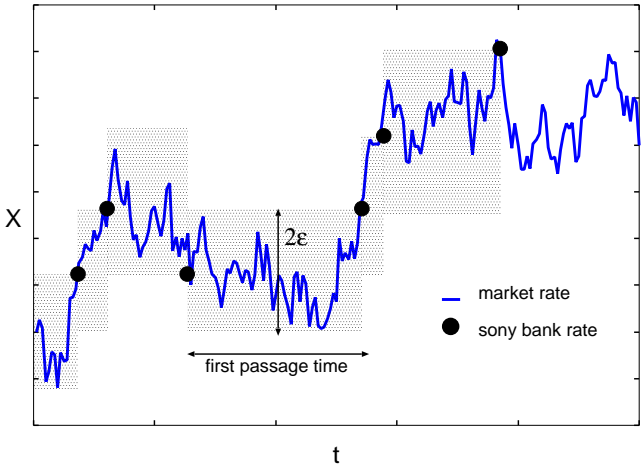}
\end{center}
\caption{\footnotesize An illustration of 
generating the filtered rate 
(black circle) by the 
rate window with width $2\epsilon$ 
(shaded area) from the 
market rate (solid line).}
\label{fig:fg_window}
\end{figure}

The Sony bank rate (see Fig. \ref{fig:fg_sony}) 
we shall deal with in this paper 
is the rate
for individual customers
of the Sony bank \cite{Sony} 
in their on-line foreign exchange trading service
via the internet.
If the USD/JPY market rate changes by 
greater or equal to $0.1$ yen, 
the Sony bank USD/JPY exchange 
rate is updated to the market rate. 
In this sense, the Sony bank rate 
can be regarded as a kind of first 
passage processes \cite{Redner,Kappen,Gardiner,Risken,Simonsen,Kurihara}. 
In Fig. \ref{fig:fg_window}, 
we show the mechanism of generating the Sony bank rate 
from the market rate (This process is sometimes 
refereed to as 
{\it first exit process} \cite{Montero}). 
As shown in the figure, 
the difference between 
successive two points in the Sony bank rate 
becomes longer than the time intervals of the market rates. 
In Table \ref{tab:table1}, 
we show 
several data concerning 
the Sony bank USD/JPY 
rate vs. tick-by-tick 
data by Bloomberg for USD/JPY rate. 
\begin{table}
\caption{\label{tab:table1}
The Sony bank USD/JPY exchange rate vs. tick-by-tick 
data for USD/JPY exchange rate. }
\begin{tabular}{|l||c|c|}
\hline
\mbox{} & Sony bank rate & tick-by-tick data \\
\hline
$\#$ of data a day & $\sim 70$  & $\sim 10,000$ \\
\hline
The smallest price change  & $0.1$ yen  & $0.01$ yen \\
\hline
Average interval between data & $\sim 20$ minutes & $\sim 7$ seconds \\
\hline
\end{tabular}
\end{table}
It is non-trivial problem to ask what kind of 
distribution is suitable to explain the distribution 
of the first passage time. 
For this problem, 
we attempted to check several statistics 
from both analytical and empirical points of view  
under the assumption that 
the first passage time might obey 
a non-exponential Weibull 
distribution \cite{Sazuka0,Sazuka2,InoueSazuka2006}. 
We found that 
the data is well explained by 
a Weibull distribution. 
This fact means that 
the difference between successive Sony bank rate changes
is fluctuated and has some memories.  
\section{Previous related results}
\label{sec:SumWorks}
Before we start to explain what we attempt to do 
in the present paper, we shortly summarize 
the related results which were already obtained by the 
present authors. 
\begin{itemize}
\item 
{\bf Sazuka (2006)}: 
We checked that the Sony bank rate is well-described 
by a Weibull distribution by some 
empirical data analysis (Weibull paper analysis, 
evaluation of 
Kullback-Leibuler, Hellinger divergence measures) 
\cite{Sazuka0,Sazuka2}. 
\item
{\bf Inoue and Sazuka (2006)}:
We showed analytically that 
the crossover between non-Gaussian 
L$\acute{\rm e}$vy regime to Gaussian regime 
is observed even in the first passage process 
(which is the same process of the Sony bank rates) 
for a truncated L$\acute{\rm e}$vy flight 
(the so-called {\it KoBoL process} 
\cite{Schoutens,Koponen,Boyarchenko} 
in mathematics or mathematical finance) 
\cite{InoueSazuka2006a}. 
\item
{\bf Inoue and Sazuka (2006)}: 
We introduced queueing theoretical approach into 
the analysis of the Sony bank rate and 
evaluated average waiting time including expected returns. 
We also carried out computer simulations by using 
the GARCH model to investigate the 
effect of the rate window of 
the Sony bank \cite{InoueSazuka2006}. 
\item
{\bf Sazuka (2006)}:
We observed a phase transition 
between a Weibull distribution to 
a power-law distribution 
at some critical time from the empirical data 
analysis of the Sony bank rate \cite{Sazuka2}.
\item 
{\bf Sazuka and Inoue (2007)}: 
We introduced the Gini index to 
evaluate to what extent a Weibull distribution 
is well-fitted to explain the behavior of the 
first passage process of the Sony bank rate. 
The analytical evaluation and the empirical data analysis 
gave quite similar results \cite{SazukaInoue2006PhysA,SazukaInoue2006}.
\end{itemize}
Then, we focus on the following 
two points. 
\begin{itemize}
\item 
The effect of discreteness of 
returns in computer simulation 
by means of the GARCH model. 
\item
Generalization of 
the renewal-reward theorem 
to calculate the higher order moments 
of the average waiting time or 
to evaluate the average waiting time 
for the case in which 
the observation time distribution of 
the trader is explicitly given.
\end{itemize}
This paper is intended as an investigation of 
these two points.  
\section{Computer simulations by 
the GARCH model with discrete returns}
\label{sec:GARCH}
In the previous studies \cite{InoueSazuka2006}, 
we carried out computer simulations of the Sony bank USD/JPY exchange 
rates by assuming that the raw market data might be 
well-described by the GARCH model \cite{Engle,Ballerslev,Franke} 
in which the successive time intervals $\Delta t$ obey a Weibull 
distribution specified by a single parameter 
$m_{0}$, namely, 
\begin{eqnarray}
P(\Delta t) & = & 
\frac{m_{0}\,(\Delta t)^{m_{0}-1}}{a}
\,{\exp}
\left[
-\frac{(\Delta t)^{m_{0}}}{a}
\right].
\end{eqnarray}
In \cite{InoueSazuka2006}, 
we set $a=1$ for a scaling parameter for simplicity 
to handle. 
We investigated the stochastic process generated 
by the rate window and estimated the distribution 
of time intervals of the stochastic process 
filtered by the rate window. 
Then, we assumed that the output of the filter 
having rate window with width $2 \epsilon$ also follows 
a Weibull distribution with 
$m (\neq m_{0})$ and estimated $m$ by 
means of Weibull paper analysis. 
The plot $m$-$m_{0}$ 
was obtained and the effect of the rate window 
on the market rate became clear. 
However, 
in those simulations, 
we assumed that the smallest price change (return) 
might take continuous values for 
simplicity to carry out the simulations. 
As we see in Table 1, 
the smallest price change unit of the 
tick-by-tick market data is $0.01$ yen. 
Therefore, 
we should modify our computer simulations 
of the GARCH model by taking into account 
the discreteness of the price change. 
In this section, we show the result 
of the computer simulations. 
To this end, 
we deal with here the following modified GARCH(1,1) model 
\cite{InoueSazuka2006}: 
\begin{eqnarray}
X_{t} & = & 
X_{t-\Delta t}+
{\mathcal N} (0,\sigma_{t}^{2}) \\
\sigma_{t}^{2} & = & 
\alpha_{0}+ 
\alpha_{1} X_{t-\Delta t}^{2} + \beta_{1}
\sigma_{t-\Delta t}^{2} 
\end{eqnarray}
where the time interval 
$\Delta t$ obeys a Weibull distribution 
with parameters $(m,a)$, namely, 
\begin{eqnarray}
P(\Delta t) & = & 
\frac{m(\Delta t)^{m-1}}{a}\,
{\exp}
\left[
-\frac{(\Delta t)^{m}}{a}
\right]
\end{eqnarray}
and ${\mathcal N} (0,\sigma_{t}^{2})$ in this expression stands for 
a Gaussian with zero-mean and 
time dependent variance $\sigma_{t}^{2}$. 
For simplicity, we set the parameter $a$ as $a=1$. 
We should notice that 
the GARCH(1,1) model 
has the variance $\sigma^{2}$ 
after observing on a long 
time intervals $t \to \infty$ and 
$\sigma^{2}$ leads to 
\begin{eqnarray}
\sigma^{2} & = & 
\frac{\alpha_{0}}
{1-\alpha_{1}-\beta_{1}}. 
\end{eqnarray}
To make the return 
$X_{t}$ discrete, 
for each time step, 
we regenerate $X_{t}$ by using 
the next map: 
\begin{eqnarray}
X_{t} & = & 
\Psi_{\Delta} (X_{t}) \equiv 
\Delta \,{\rm ceil}
(\Delta^{-1} X_{t})
\label{eq:Map}
\end{eqnarray}
where the function 
${\rm ceil}(x)$ is 
defined as 
the smallest integer 
no less than $x$. 
The parameter 
$\Delta$ appearing in 
(\ref{eq:Map}) means 
the length of 
the minimal variation 
of the return. 
\begin{figure}[ht]
\begin{center}
\rotatebox{-90}{\includegraphics[width=6.5cm]{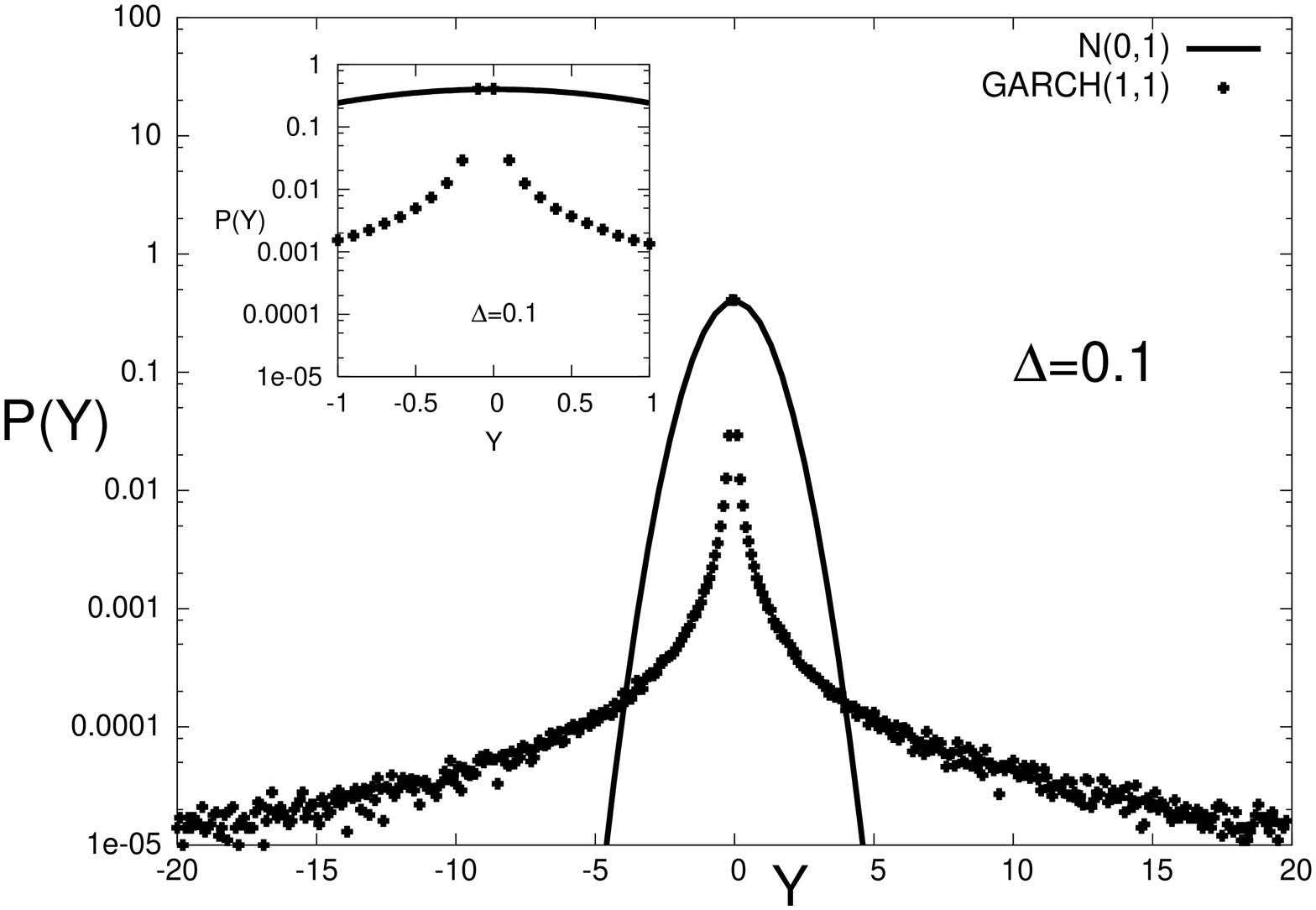}} \hspace{0.3cm}
\rotatebox{-90}{\includegraphics[width=6.5cm]{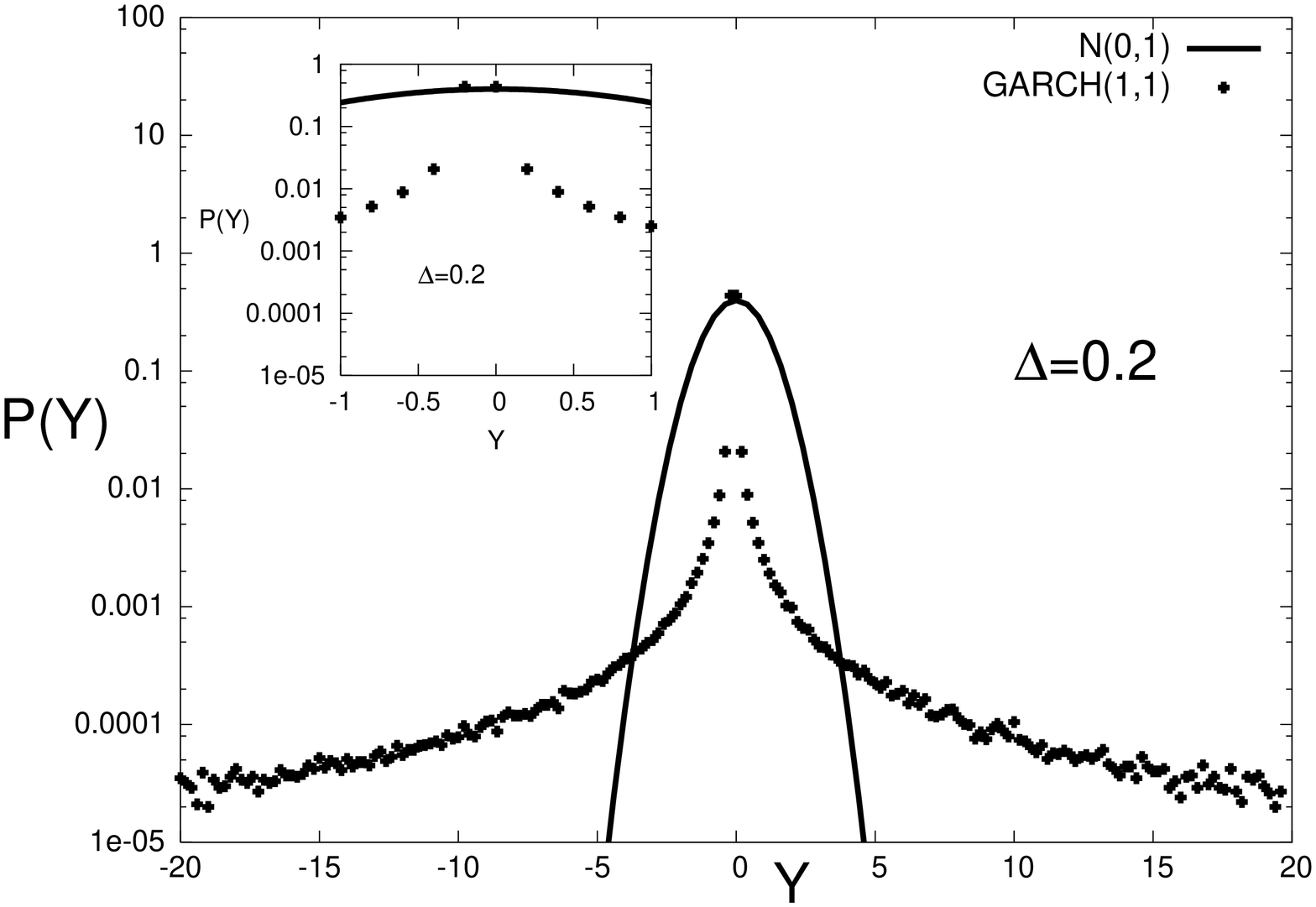}}
\end{center}
\caption{\footnotesize 
The pdf of the GARCH(1,1) model with discrete returns. 
We set $\Delta=0.1$ (upper panel) and 
$\Delta=0.2$ (lower panel). 
The inset is behavior around small $Y$ regime.}
\label{fig:fg7}
\end{figure} 
In Fig. \ref{fig:fg7}, 
we plot the pdf of 
$Y_{t}=X_{t}-X_{t-\Delta t}$ for 
the above discrete GARCH model with a parameter set: 
$(\alpha_{0},\alpha_{1},\beta_{1})=(0.4,0.3,0.3)$ 
which gives 
the kurtosis $\kappa = 4.17$ and 
variance $\sigma^{2}=1$ in 
the limit of $t \to \infty$, and 
$\Delta=0.1$ and $\epsilon=\sigma=1$. 
We should keep in mind that the ratio 
between 
the width of the rate window and 
the minimal change of the rate, namely, 
$\epsilon/\Delta =10$ is 
the same as that of the Sony bank rate.  
From this figure, 
we recognize that the return $Y$ actually takes 
discrete values as expected. 

To investigate 
the effect of the rate window 
with width $2 \epsilon$, 
we estimate 
the Weibull parameter $m$ 
of the output sequence from 
the rate window by means of the so-called 
{\it Weibull paper analysis} 
\cite{Sazuka0,Sazuka2,InoueSazuka2006} 
for the cumulative Weibull distribution. 
\begin{figure}[ht]
\begin{center}
\rotatebox{-90}{\includegraphics[width=6.5cm]{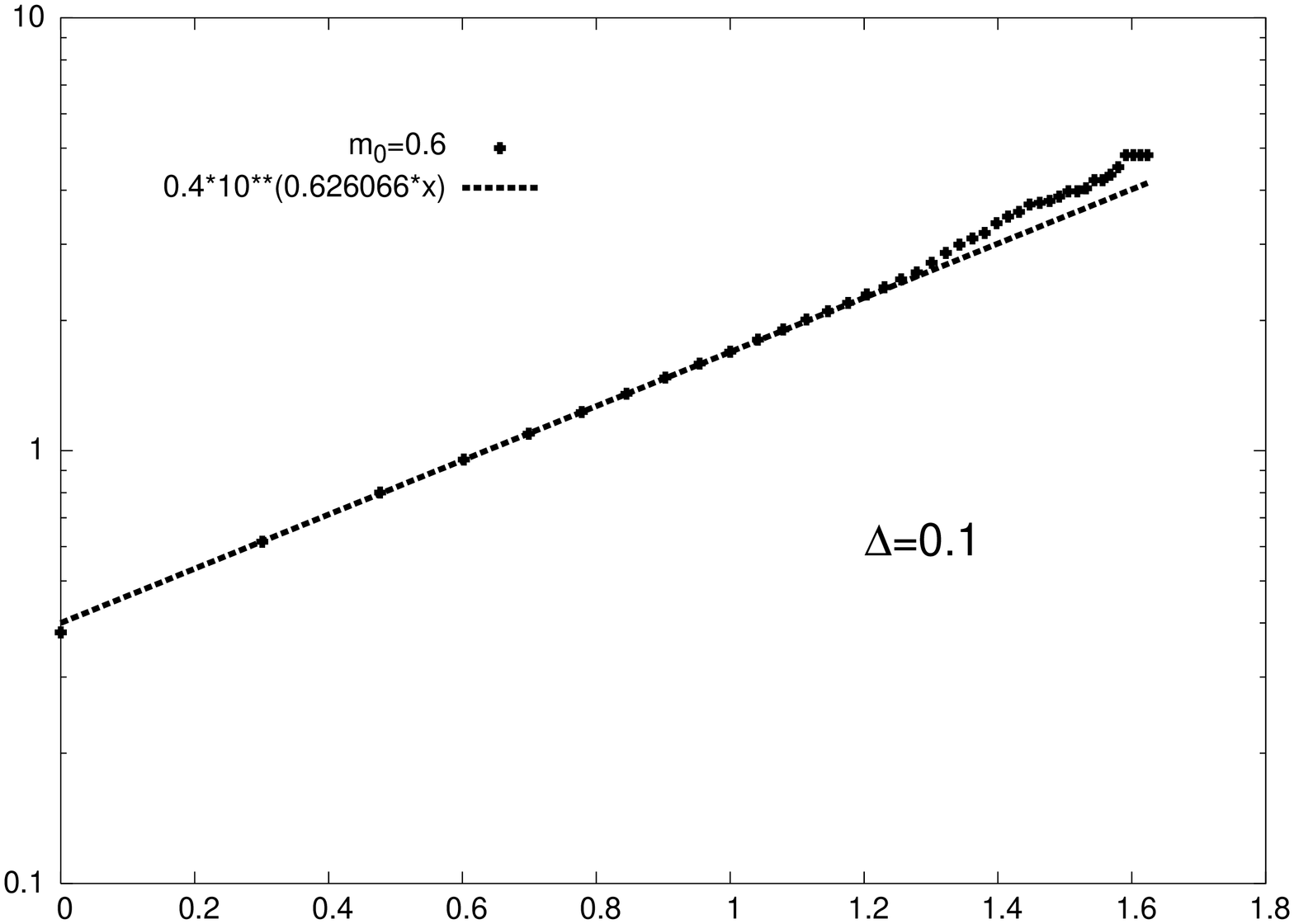}} \hspace{0.3cm}
\rotatebox{-90}{\includegraphics[width=6.5cm]{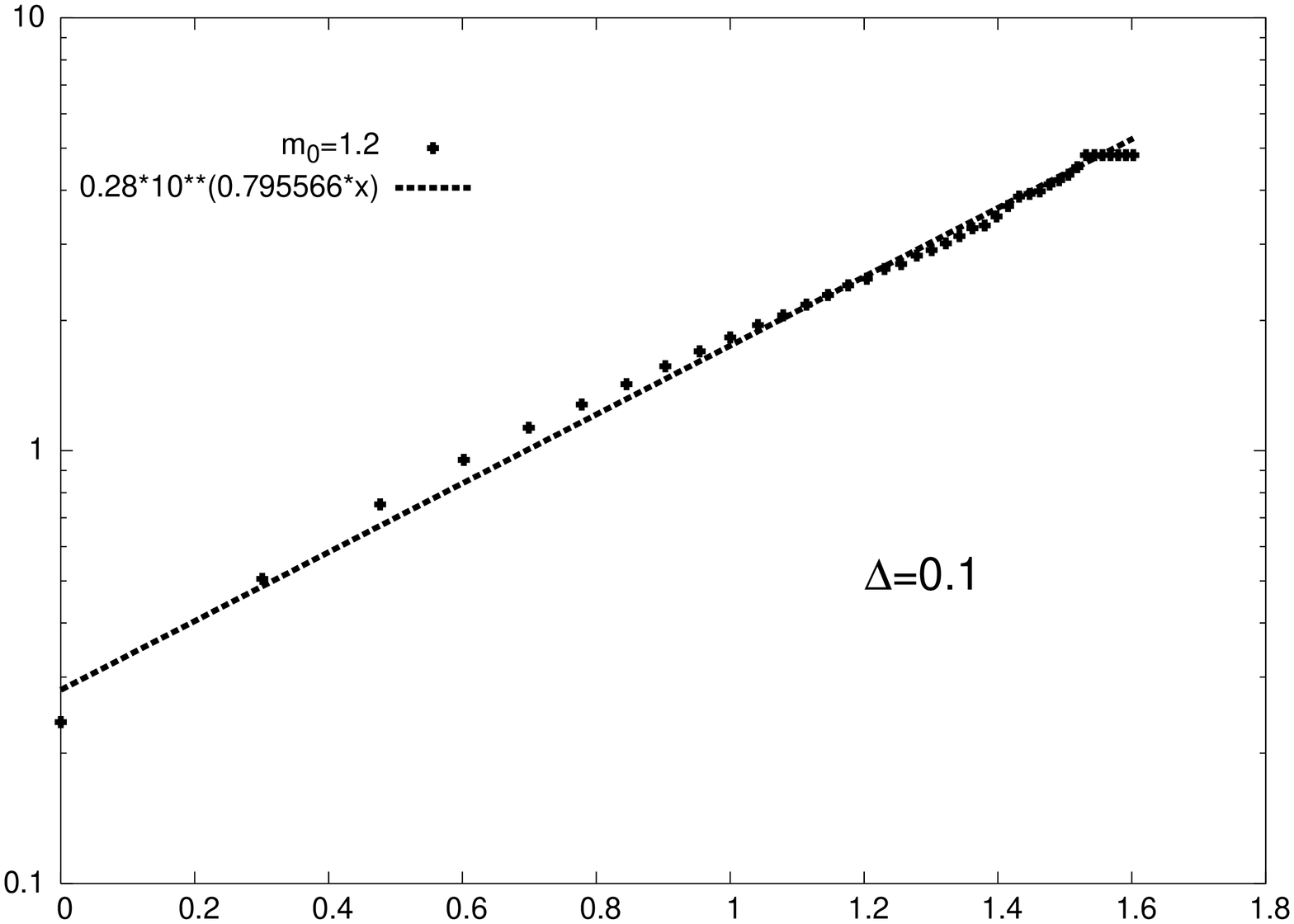}}
\end{center}
\caption{\footnotesize 
The Weibull paper for 
the case of $m_{0}=0.6$ (upper panel) and 
$1.2$ (lower panel). 
We choose 
$\Delta=0.1$. }
\label{fig:fg8}
\end{figure} 
From 
Fig. \ref{fig:fg8}, 
we find that for 
the case of $m_{0}=0.6,1.2$, 
the first passage time distribution 
also obeys a Weibull distribution 
with a parameter $m$ which is different from 
$m_{0}$. 
\begin{figure}[ht]
\begin{center}
\rotatebox{-90}{\includegraphics[width=6.5cm]{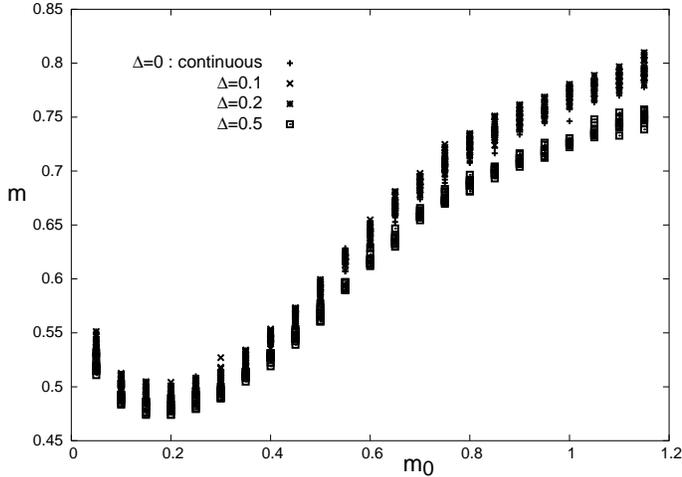}}
\end{center}
\caption{\footnotesize 
The relation between 
$m_{0}$ and $m$ for 
the case of $\Delta=0,0.1,0.2$ and $0.5$. 
For each $\Delta$, 
we carried out $10$ independent runs 
of simulations.} 
\label{fig:fg9}
\end{figure} 
In Fig. \ref{fig:fg9}, 
we plot the relation 
between $m_{0}$ and $m$ for 
several values of 
$\Delta$. 
From this figure, we
find that the relation for the 
discrete cases 
with $\Delta =0.1$ and $0.2$ are 
almost same as the 
relation for 
the continuous case 
$\Delta=0$ reported in our previous papers \cite{InoueSazuka2006}. 
Meaningful differences are 
observed if we increase 
the value of $\Delta$ up to $0.5$. 
This result provides us a justification 
of our previous GARCH modeling 
\cite{InoueSazuka2006} of the market rate 
to simulate the Sony bank USD/JPY exchange rates 
on the assumption that the minimum price change 
could take infinitesimal values.

\section{Derivation of the waiting time distribution}
\label{sec:Deriv}

In the previous studies \cite{InoueSazuka2006}, 
we evaluated the average waiting time 
for the customers to wait by the next price change since 
they attempt to observe the price by making use of 
the renewal-reward theorem \cite{Tijms,Oishi}. 
However, the theorem itself is obviously restricted to 
deriving only the first moment of the waiting time. 
From this reason, it is very hard to 
evaluate, for instance, 
the standard deviation from the average waiting time 
within the framework of the theorem 
(for example, see the proof of 
the theorem provided in \cite{Oishi}). 
Thus, we need another procedure to calculate it 
without the conventional derivation of the 
renewal-reward theorem. 
In this section, we directly derive the distribution 
of the waiting time. 
Our approach here enables us to 
evaluate not only the first moment of the 
waiting time but also any order of the moment. 
This section is a core part of this paper. 

\subsection{The probability distribution of the waiting time}
\begin{figure}[ht]
\begin{center}
\includegraphics[width=8.5cm]{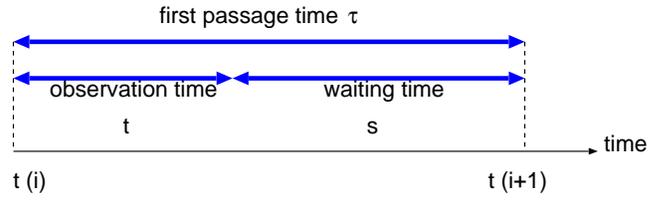}
\end{center}
\caption{\footnotesize 
The relation 
these points 
$\tau,t$ and $s$
in time axis. 
The first passage time $\tau$ 
is given by 
$\tau=t(i+1)-t(i)$. 
The observation time is measured from 
the point $t (i)$. 
}
\label{fig:fg9b}
\end{figure}

We first derive the probability distribution function of 
the waiting time $s$. 
Then, let us suppose that 
the difference between 
successive two points 
of the Sony bank rate change, 
namely, the first passage time 
$\tau$ 
follows $P_{W}(\tau)$. 
Then, the customers observe the 
rate in time $t$ 
($0 \leq t \leq \tau$) 
that should be 
measured from the point at which 
the rate changes previously. 
In Fig. \ref{fig:fg9b}, 
we show the relation 
among these points 
$\tau,t$ and $s$
in time axis. 
The waiting time of 
the customers 
is naturally defined by 
$s \equiv \tau - t$. 
We should notice that 
the distribution $\Omega (s)$ 
is written in terms of the first passage time 
distribution $P_{W}(\tau)$ and the observation time distribution 
$P_{O}$(t) of the customers as 
\begin{eqnarray}
\Omega (s) & \propto & 
\int_{0}^{\infty}
d\tau 
\int_{0}^{\tau}
dt 
\,Q(s|\tau,t)P_{O}(t)P_{W}(\tau). 
\end{eqnarray}
Obviously,   
probability 
$Q(s|\tau,t)$ that 
the waiting time 
takes $s$ provided that 
the observation time and 
the first passage time 
are given as $t$ and $\tau$, 
respectively, 
is given as 
\begin{eqnarray}
Q(s|\tau,t) & = & 
\delta (s-\tau+t)
\end{eqnarray}
with the delta function 
$\delta (x)$. 
Taking into account the 
normalization constant of 
$\Omega (s)$, we have
\begin{equation}
\Omega (s) =
\frac{\int_{0}^{\infty}
d\tau P_{W}(\tau) 
\int_{0}^{\tau}
dt \,\delta (s-\tau +t) P_{O}(t)} 
{
\int_{0}^{\infty}
ds
\int_{0}^{\infty}
d\tau P_{W}(\tau) 
\int_{0}^{\tau} 
dt \,\delta (s-\tau + t)P_{O}(t)} 
\end{equation}
where $t$ denotes the observation time for the customers. 
We should notice that the result of the renewal-reward theorem : 
$w=\langle s \rangle = E(\tau^{2})/2E(\tau)$ 
(see for example \cite{Oishi}) is 
recovered by inserting the uniform observation time 
distribution $P_{O}(t)=1$ 
into the above expression as 
\begin{eqnarray}
w & = & 
\langle s \rangle = 
\int_{0}^{\infty}
ds s \Omega (s) = 
\frac{
\int_{0}^{\infty}
ds s 
\int_{s}^{\infty}
d\tau 
P_{W}(\tau)}
{
\int_{0}^{\infty}
ds 
\int_{s}^{\infty}
d\tau 
P_{W}(\tau)} \nonumber \\
\mbox{} & = & 
\frac{
\int_{0}^{\infty}
\frac{d}{dt}\{s^{2}/2\} ds
\int_{s}^{\infty}
d\tau 
P_{W}(\tau)}
{
\int_{0}^{\infty}
\frac{d}{ds}\{s\} ds 
\int_{s}^{\infty}
d\tau 
P_{W}(\tau)} \nonumber \\
\mbox{} & = & 
\frac{
(1/2) \int_{0}^{\infty}
s^{2}P_{W}(s)ds}
{
\int_{0}^{\infty}
sP_{W}(s)ds} = 
\frac{E(\tau^{2})}{2E(\tau)}
\end{eqnarray}
where we defined 
the $n$-th moment of the 
first passage time 
$E(\tau^{n})$ by 
\begin{eqnarray}
E(\tau^{n}) & = & 
\int_{0}^{\infty}
ds s^{n} 
P_{W}(s).
\end{eqnarray}
More generally, we may set 
$P_{O}(t)$. 
For this general form of 
the observation time distribution, 
the probability distribution of 
the waiting time $s$ is given as follows. 
\begin{eqnarray}
\Omega (s) & = & 
\frac{
\int_{s}^{\infty}
d\tau 
P_{W}(\tau) P_{O} (\tau-s)}
{
\int_{0}^{\infty}ds
\int_{s}^{\infty}
d\tau 
P_{W}(\tau) P_{O}(\tau-s)} \nonumber \\
\mbox{} & = &
\frac{
\int_{s}^{\infty}
d\tau 
P_{W}(\tau) P_{O} (\tau-s)}
{E(t) - \delta_{1}}
\label{eq:Omega_s}
\end{eqnarray}
where we defined $\delta_{n}$ by 
\begin{eqnarray}
\delta_{n} & = & 
\int_{0}^{\infty}
\frac{ds s^{n}}{n}
\int_{s}^{\infty}
P_{W}(\tau) 
\frac{\partial P_{O} (\tau-s)}
{\partial s}.
\end{eqnarray}
By using the same way as the derivation of the 
distribution 
$\Omega (s)$, 
we easily obtained 
the first two moments of 
the waiting time distribution as
\begin{eqnarray}
\langle s \rangle & = & 
\frac{E(\tau^{2})/2 - 
\delta_{2}}
{E(\tau)-\delta_{1}},\,\,\,
\langle s^{2} \rangle = 
\frac{E(\tau^{3})/3- 
\delta_{3}}
{E(\tau)-\delta_{1}}
\label{eq:two_Moments}
\end{eqnarray}
and the standard deviation leads to 
\begin{equation}
\sigma = 
\sqrt{
\frac{
\{
4E(\tau^{3})E(\tau) - 
3E(\tau^{2})\}
+G_{\delta_{1},\delta_{2},\delta_{3}}  
}
{
12(E(\tau)-\delta_{1})^{2}
}
}
\label{eq:Deviation}
\end{equation}
\begin{eqnarray}
G_{\delta_{1},\delta_{2},\delta_{3}} & = & 
- 
4\delta_{1} E(\tau^{3}) - 
12\delta_{3}E(\tau)  \nonumber \\
\mbox{} & + & 
12\delta_{2}E(\tau^{2}) + 12\delta_{1} \delta_{3} 
-12\delta_{2}^{2}
\end{eqnarray}
where 
we defined 
\begin{eqnarray}
\langle s^{n} \rangle & = & 
\int_{0}^{\infty}
dss^{n}\Omega (s).
\end{eqnarray}
Thus, this probability distribution $\Omega (s)$ enables us 
to evaluate any order of the moments for the waiting time. 

In following, we evaluate the average waiting time 
and the deviation around 
the average for typical two cases, 
namely, 
$P_{O}(t)=1$ and 
$P_{O}(t)={\rm e}^{-t/\tau_{0}}$. 
\subsubsection{The customers' observation 
follows $P_{O} (t)=1$}

\begin{figure}[ht]
\begin{center}
\rotatebox{-90}{\includegraphics[width=6.5cm]{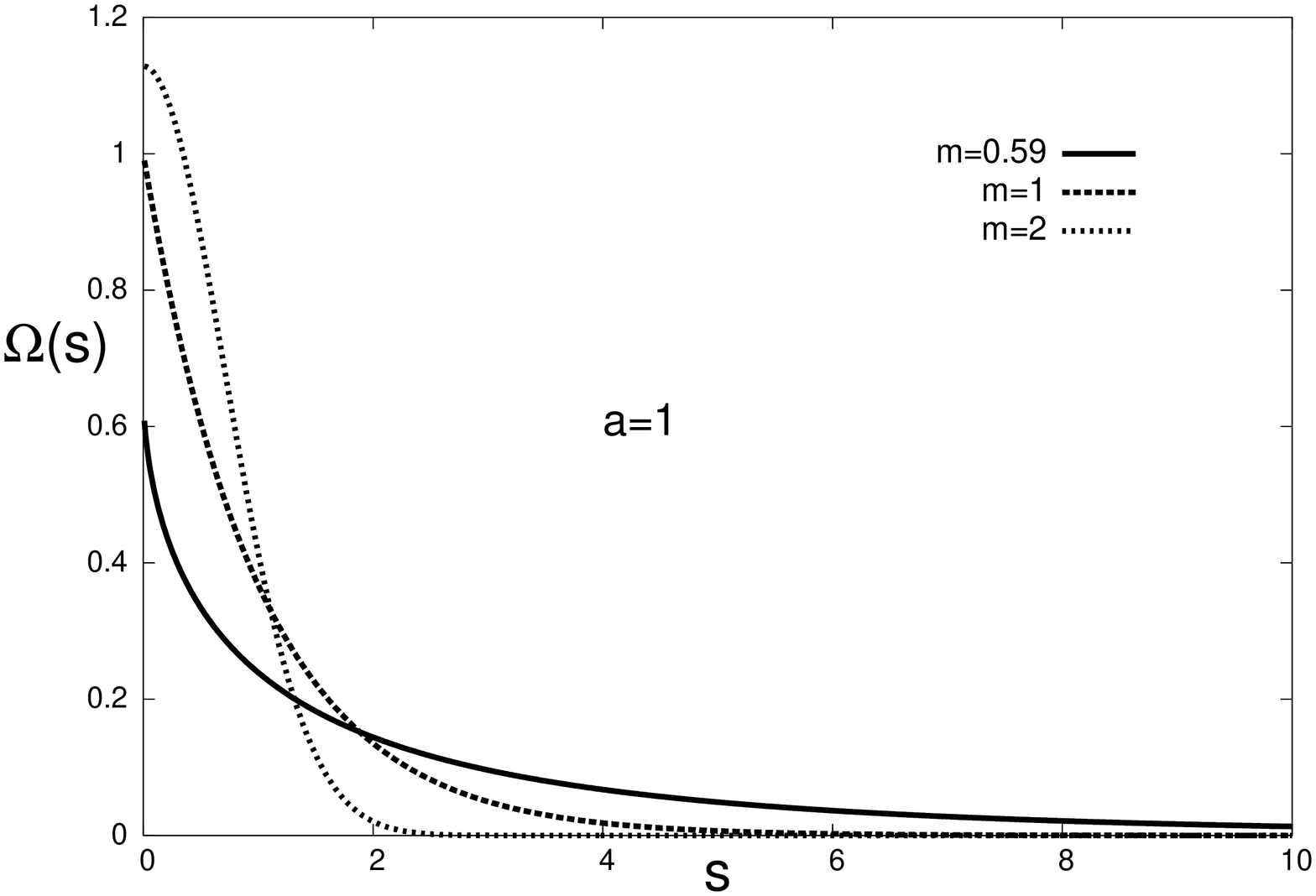}} \hspace{0.3cm}
\rotatebox{-90}{\includegraphics[width=6.5cm]{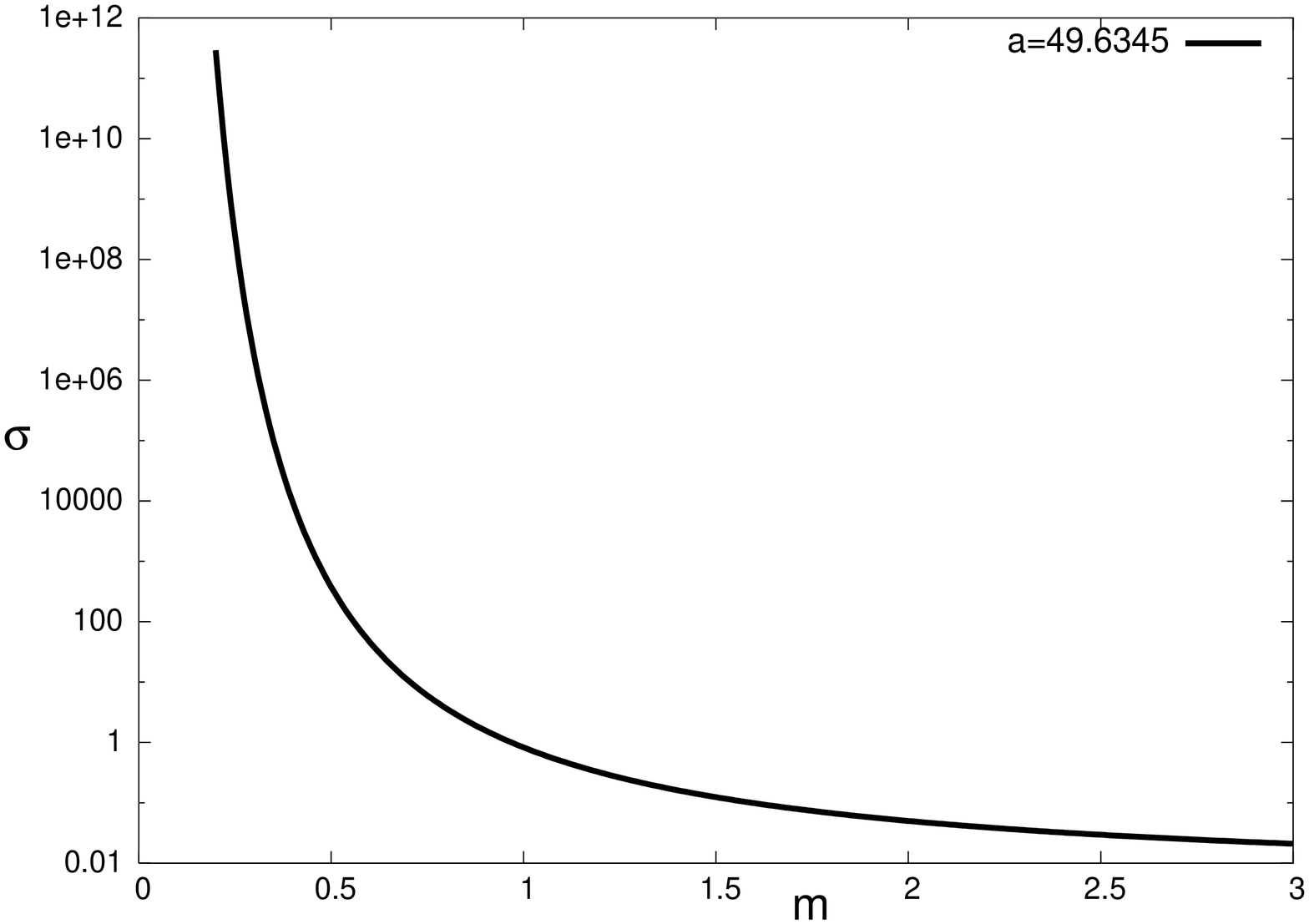}}
\end{center}
\caption{\footnotesize 
The distribution 
of waiting time 
for a Weibull distributing 
$\Omega (s)$ with $a=1$ and 
$m=0.59,1$ and $2$ (upper panel). 
In the lower panel, we 
plot the standard deviation 
$\sigma$ for a Weibull 
distribution with $a=49.6345$ as 
a function of $m$. }
\label{fig:fg9c}
\end{figure}

We first consider the case of 
$P_{O}(\tau)=1$. 
This case corresponds to 
the result obtained by the renewal-reward theorem 
\cite{InoueSazuka2006}. 
Obviously, we find that 
$\delta_{n} =0$ holds for 
arbitrary integer $n$. 
Thus, 
the waiting time distribution 
$\Omega (s)$ leads to 
\begin{eqnarray}
\Omega (s) & = & 
\frac{\int_{s}^{\infty}
P_{W}(\tau)}
{E(\tau)}. 
\end{eqnarray}
Then, the average waiting time and 
the deviation around the value lead to 
\begin{eqnarray}
w & = & 
\frac{E(\tau^{2})}{2E(\tau)},\,\,\,\,
\sigma = 
\sqrt{
\frac{4E(\tau^{3})E(\tau) - 
3E(\tau^{2})^{2}}
{12E(\tau)^{2}}
}.
\label{eq:def_dev}
\end{eqnarray}
For a Weibull distribution 
having the parameters $m,a$, 
the above results are rewritten by 
\begin{eqnarray}
\Omega (s) & = & 
\frac{m \, {\rm e}^{-s^{m}/a}}
{a^{1/m} \Gamma 
\left(
\frac{1}{m}
\right)} \\
w & = & 
a^{1/m}
\frac{\Gamma 
\left(
\frac{2}{m}
\right)}
{
\Gamma 
\left(
\frac{1}{m}
\right)
} \\
\sigma & = & 
\frac{a^{1/m} \sqrt{
\Gamma(1/m)\Gamma (3/m) - 
\Gamma (2/m)^{2}}}
{\Gamma (1/m)}
\end{eqnarray}
where we defined the Gamma 
function as
\begin{eqnarray}
\Gamma(x) & = & 
\int_{0}^{\infty}dt\, 
t^{x-1}{\rm e}^{-t}.
\end{eqnarray}
It is important for us 
to notice that 
for an exponential 
distribution 
$m=1$, 
we have 
$w=\sigma=a$ 
by taking into account the fact that 
$\Gamma (n)=(n-1)!$. 
Moreover,
the average waiting time $w$
is identical to the average time interval
$E(\tau)$
since 
$w=E(\tau^{2})/2E(\tau)=E(\tau)$ 
holds 
if and only if 
$m=1$ (The rate changes follow a Poisson 
arrival process).
These results are 
already obtained in 
our previous studies \cite{InoueSazuka2006}. 
In 
Fig. \ref{fig:fg9c}, 
we plot the 
distribution 
$\Omega (s)$ and 
the standard deviation 
$\sigma$ for a Weibull 
distribution. 
Especially, for 
the Sony bank case 
$a=49.6345,m=0.585$, 
we find 
$\sigma=60.2284$ minutes. 
\subsubsection{The customers' observation follows 
$P_{O} (t) = {\rm e}^{-t/\tau_{0}}$}
We test the other observation time distribution of 
the customers. For instance, we might choose 
$P_{O} (t)= {\rm e}^{-t/\tau_{0}}$ 
on the assumption that 
the customers might observe the rate 
more frequently within the time scale 
$\tau_{0}$ around 
the previous rate change 
than the time scale longer than $\tau_{0}$. 
Such a case could be 
possible if the system 
provides some redundant information about the 
point of the rate change to customers 
as quickly as possible. 

\begin{figure}[ht]
\begin{center}
\rotatebox{-90}{\includegraphics[width=6.5cm]{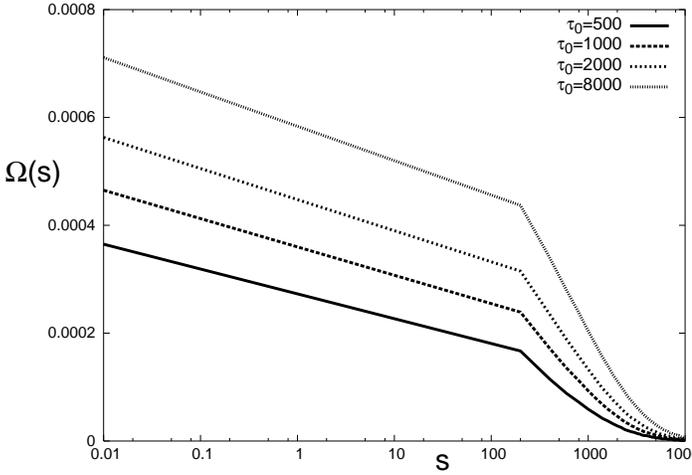}}
\end{center}
\caption{\footnotesize 
The waiting time distribution 
$\Omega (s)$ for 
the case of the weight function 
$P_{O} (t)={\rm e}^{-t/\tau_{0}}$ 
with several values of $\tau_{0}$. 
For each plot, we set 
$m=0.585$ and $a=49.6345$. }
\label{fig:fg10}
\end{figure}

For this choice of the observation time distribution 
$P_{O} (t)=
{\rm e}^{-t/\tau_{0}}$, we have
\begin{eqnarray}
\delta_{n} & = & 
\frac{1}{\tau_{0}}
\int_{0}^{\infty}
\frac{ds s^{n} \, {\rm e}^{s/\tau_{0}}}
{n}
\int_{s}^{\infty}
d\tau 
P_{W}(\tau)\, 
{\rm e}^{-\tau/\tau_{0}}.
\end{eqnarray}
Therefore, 
from equations (\ref{eq:Omega_s}),(\ref{eq:two_Moments}) 
and (\ref{eq:Deviation}), 
it is possible for us to derive 
the waiting time distribution 
$\Omega (s)$ and 
average waiting time and 
the deviation from 
the value. 
For the choice of 
a Weibull first passage time 
distribution $P_{W}(\tau)$, 
we need some algebra, but easily obtain 
\begin{eqnarray}
\Omega (s) & = &  
\frac{
{\rm e}^{s/\tau_{0}}
\mu_{a}^{m}
(\tau_{0};s)}
{\frac{a^{1/m}}{m}
\Gamma 
\left(
\frac{1}{m}
\right) - 
\frac{1}{\tau_{0}}
\int_{0}^{\infty}
sds \,
\mu_{a}^{m} (\tau_{0};s)
}
\end{eqnarray}
where we defined $\mu_{a}^{m}(\tau_{0};s)$ by 
the following integral form: 
\begin{eqnarray}
\mu_{a}^{m} 
(\tau_{0};s) & = & 
\int_{a^{-1}s^{m}}^{\infty}
dz \, 
{\exp}
\left[
-z -
\frac{(az)^{1/m}}
{\tau_{0}}
\right].
\end{eqnarray}
In Fig. \ref{fig:fg10}, 
we plot the distribution 
for the case of several values of $\tau_{0}$ and 
$m=0.585$ and $a=49.6345$. 
From this figure, we find that 
the probability for large waiting time 
$s$ increases as the relaxation 
time $\tau_{0}$ increases. 
We should notice that the case of 
$\tau_{0} \to \infty$ turns out to be 
random observation by the customers. 

From the equations (\ref{eq:two_Moments}), 
the first two moments are obtained by 
\begin{eqnarray}
w & = & 
\langle s \rangle = 
\frac{
\frac{a^{2/m}}{m}
\Gamma 
\left(
\frac{2}{m}
\right) - 
\tau_{0}^{-1}
\xi_{m,a}^{(2)}(\tau_{0})}
{
\frac{a^{1/m}}{m}
\Gamma
\left(
\frac{1}{m}
\right) - 
\tau_{0}^{-1}
\xi_{m,a}^{(1)}
(\tau_{0})} \\
\langle s^{2} \rangle & = & 
\frac{
\frac{a^{3/m}}{m}
\Gamma 
\left(
\frac{3}{m}
\right) - 
\tau_{0}^{-1}
\xi_{m,a}^{(3)}(\tau_{0})}
{
\frac{a^{1/m}}{m}
\Gamma
\left(
\frac{1}{m}
\right) - 
\tau_{0}^{-1}
\xi_{m,a}^{(1)}
(\tau_{0})}
\end{eqnarray}
where we defined
\begin{eqnarray}
\xi_{m,a}^{(n)}(\tau_{0}) & = & 
\int_{0}^{\infty}
\frac{ds s^{n} {\rm e}^{s/\tau_{0}}}
{n}
\mu_{a}^{m}
(\tau_{0};s).
\end{eqnarray}
It should be noted that 
for large $\tau_{0}$, 
the leading order of 
the function $\xi_{m,a}^{(n)}(\tau)$ 
is evaluated as 
$\xi_{m,a}^{(n)}(\tau) = 
1 + {\mathcal O}(\tau_{0}^{-1})$. 
Thus, the average waiting time $w$
for $P_{O} (t)={\rm e}^{-t/\tau_{0}}$
becomes smaller than that obtained 
for $P_{O} (t)=1$ as 
\begin{eqnarray}
w & = & 
a^{1/m} 
\frac{
\Gamma 
\left(
\frac{2}{m}
\right)}
{
\Gamma 
\left(
\frac{1}{m}
\right)
} - 
{\mathcal O}
(\tau_{0}^{-1}).
\end{eqnarray}
However, at the same time, the standard deviation also 
behaves as 
\begin{equation}
\sigma = 
\frac{a^{1/m} 
\sqrt{
\Gamma (1/m)
\Gamma (3/m) - 
\Gamma (2/m)^{2}
}
}
{\Gamma (1/m)}
 - 
{\mathcal O}(\tau_{0}^{-1}). 
\end{equation}
This means that 
even if the trader observe 
the rate according to 
{\it a priori} knowledge, 
namely, the observation time distribution 
$P_{O}(t)={\rm e}^{-t/\tau_{0}}$, 
the standard deviation from 
the average waiting time 
is the same order as the average waiting time itself. 

In this paper, 
we considered just only two cases 
of the observation time distribution $P_{O}(t)$, however, 
the choice of the distribution 
is completely arbitrary. 
The detail and more carefully analysis 
of this point will be reported in 
our forth coming article \cite{SazukaInoue2006}. 
\subsubsection{Comparison with empirical data analysis}
It is time for us to compare the analytical result with 
that of the empirical data analysis. 
For uniform observation 
time distribution 
$P_{O} (t)=1$, 
we obtained 
$\sigma=60.2284$ minutes. 
On the other hand, 
from empirical data analysis, we 
evaluate 
the 
quantity (\ref{eq:def_dev}) 
by sampling 
the moment 
as $E(\tau^{n}) = 
(1/N)\sum_{i=1}^{N}
\tau_{i}^{n}$ directly 
from Sony bank rate data \cite{Sony} 
and find $\sigma = 74.3464$ minutes. 
There exists a finite gap between 
the theoretical prediction and the result by 
the empirical data analysis, however, 
the both are the same order. 
The gap might become small if we 
take into account the power-law tail 
of the first passage time distribution. 
In fact, our preliminary 
investigation shows that 
for the average waiting time, 
the power-law tail makes 
the gap between the theoretical prediction and 
empirical observation small \cite{SazukaInoue2006}. 
Therefore, the same tail-effect is reasonably expected even in 
the analysis of the deviation. 
\section{Concluding remarks}
\label{sec:Summary}

In this paper, we proposed a different procedure 
from the conventional derivation of the 
renewal-reward theorem. 
This derivation enables us 
to evaluate arbitrary order of the moments of the waiting time 
of the on-line foreign exchange trading rate 
for the individual customers. 
We directly derived the waiting time 
distribution and evaluated the deviation 
around the average waiting time, 
which is not supported by the renewal-reward theorem, 
for the Sony bank USD/JPY exchange rates. 
We tested our analysis 
for several cases of the observation time distribution 
of the customers and 
found that the average waiting time 
and deviation from the value are the same order 
even if the customers 
possesses {\it a priori} knowledge as 
a form of the observation time 
distribution as $P_{O}(t)={\rm e}^{-t/\tau_{0}}$. 
This result might be understood as follows. 
As we mentioned, the system we dealt 
with in this paper has 
two types of 
fluctuations, 
namely, 
fluctuation in 
the intervals of events (price changes) and 
fluctuation in the observation time 
for the individual customers to observe the rate 
through the World Wide Web. 
As well-known, 
in the $N$-body systems ($N$ is extremely large), 
there exists theremal fluctuation 
(quantum-mechanical fluctuation as well in low temperature) 
affecting on each particle, however, the macroscopic 
variables (the averages over the Gibbs distribution) 
like pressure or internal energy 
are determined as of order $1$ objects and 
the deviation from the average becomes zero as 
$N^{-1/2}$. 
This is a reason why 
statistical mechanics 
could predict a lot of 
physical quantities and 
we could compare the prediction 
with the same quantity which was 
observed in experiments. 
On the other hand, 
although 
the financial market price changes are 
taken place as a result of 
trading by many people, 
the observed rate itself is regarded as 
a result of effective single 
particle problems (a complicated single 
random walker) and 
there is no such scale parameter like the 
number of particle $N$. 
This is an intuitive account of 
the result, namely, the average and 
the deviation are the same order 
even if there exist two kinds of 
fluctuation in the systems. 

Remaining problems concerning 
the analysis of the Sony bank rate 
are firstly to investigate the effect of the tail 
of the first passage time distribution. 
Our preliminary 
observation from the empirical data 
implied that, 
at some critical point, 
the distribution 
changes its shape 
from a Weibull-law to a power-law \cite{Sazuka2}. 
Therefore, 
it is important for us 
to check to what extent 
the prediction by the 
renewal-reward theorem 
is modified by 
taking into account the tail 
effect of 
the first passage time distribution. 
Secondly, 
we should show explicitly that the 
first passage time distribution described by the 
Mittag-Leffler function \cite{Raberto,Scalas,Gorenflo}: 
\begin{eqnarray}
P(\tau) & = & 
\sum_{n=0}^{\infty}
(-1)^{n}
\frac{(\tau/\tau_{0})^{\beta n}}
{\Gamma (\beta n +1)}\,\,\,\,\,\,\,
(0 < \beta <1)
\end{eqnarray}
is impossible to 
realize the first passage process 
of the Sony bank 
by comparing the 
analytical result of 
the average waiting time or 
the Gini index with 
those obtained by the empirical data analysis. 
The detail of these studies will be reported 
shortly in \cite{SazukaInoue2006}. 

As we showed in this paper, 
our queueing theoretical 
approach might be useful for us to build 
artificial markets such as the on-line trading 
service so as to have a suitable 
waiting time for the individual customers 
by controlling the width of the rate window.  
Moreover, theoretical framework we provided here 
could predict the average waiting time including 
the deviation. 

We hope that our strategy in order to 
analyze the stochastic process of markets 
from the view point of the waiting time 
of the customers might help 
researchers or engineers when they attempt to construct 
a suitable system for their customers.

\section*{Acknowledgment}
\label{sec:Ackno}
One of the authors (J.I.) 
was financially supported 
by {\it Grant-in-Aid 
Scientific Research on Priority Areas 
``Deepening and Expansion of Statistical Mechanical Informatics (DEX-SMI)" 
of The Ministry of Education, Culture, 
Sports, Science and Technology (MEXT)} 
No. 18079001. 
N.S. would like to appreciate 
Shigeru Ishi, President of the Sony bank,
for kindly providing the Sony bank data
and useful discussions.
We gratefully thank Enrico Scalas for fruitful discussion and 
variable comments. 

\end{document}